\documentclass[prl,amsmath,amssymb,twocolumn]{revtex4}
\usepackage{graphicx}
\usepackage{dcolumn}
\usepackage{bm}

\begin{document}

\title{Quantitative analysis of electronic transport through weakly-coupled metal/organic interfaces}

\author{A.S. Molinari$^{a}$, I. Guti\'{e}rrez Lezama$^{a}$, P. Parisse$^{a,b}$, T. Takenobu$^{c}$, Y. Iwasa$^{c}$ and A.F. Morpurgo$^{a}$}
\affiliation{$^{a}$Kavli Institute of Nanoscience, Delft University of Technology, Lorentzweg 1, 2628CJ Delft, The Netherlands\\
$^{b}$ CASTI CNR-INFM regional Lab and Dipartimento di Fisica, Universit$\grave{a}$ degli Studi dell'Aquila, Italy\\
$^{c}$ Institute for Material Research, Tohoku University, Sendai, Japan and CREST; Japan Science and Technology Corporation, Kawaguchi, Japan}

\date{\today}

\begin{abstract}
Using single-crystal transistors, we have performed a systematic experimental study of electronic transport through oxidized copper/rubrene interfaces as a function of temperature and bias. We find that the measurements can be reproduced quantitatively in terms of the thermionic emission theory for Schottky diodes, if the effect of the bias-induced barrier lowering is included.
Our analysis emphasizes the role of the coupling between metal and molecules, which in our devices is weak due to the presence of an oxide layer at the
surface of the copper electrodes.
\end{abstract}

\maketitle

\newpage
The electronic transport properties of contacts between metals and semiconductors are notoriously sensitive to the microscopic details of the interface between the two materials. For inorganic semiconductors such as Silicon, the level of material control is sufficient for the quantitative analysis of transport in terms of established theoretical models based on the concept of Schottky barrier \cite{Sze}. For organic semiconductors, on the contrary, the control of the interfacial properties (e.g., disorder present in the molecular material, the electronic coupling between metal and molecules, the density of surface states, etc.) is poor and no systematic quantitative comparison between experiments and theory has been possible so far \cite{Mall}. As a consequence, our understanding of transport across metal/organic interfaces is limited and it is not even known whether the conventional Schottky theory developed for inorganic semiconductors works also for organic materials.\\
Recently, we have demonstrated that the electrical characteristics of oxidized Cu, Ni, and Co contacts in organic single-crystal field-effect transistors (FETs) exhibit an excellent level of reproducibility \cite{APL_repr}. Here, we exploit this reproducibility to perform a systematic study of bias and temperature dependent transport across oxidized copper/rubrene interfaces. We find that the electrical characteristics of these contacts are well described quantitatively by the conventional Schottky theory with meaningful values for the microscopic parameters, if the bias-induced lowering of the Schottky barrier is taken into account. We attribute the good agreement -as well as the experimental reproducibility- to the presence of the oxide layer at the surface of copper, which causes the coupling between metal and molecules to be weak.\\ 
\begin{figure}[ht]
\centering
\includegraphics[width=1\columnwidth]{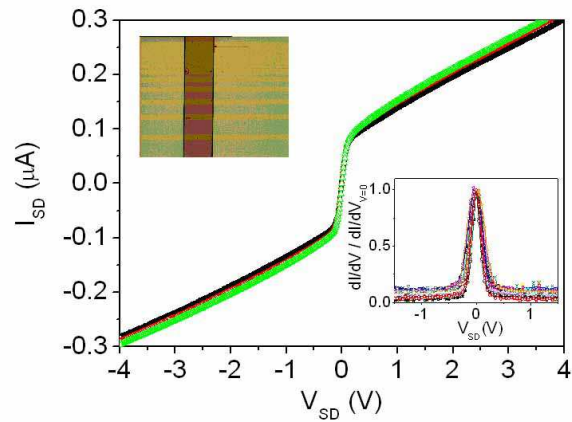}
\noindent{\caption{$I-V$ curves measured at room temperature on a short channel FET for three different gate voltages ($V_{G}=-10,-20,-30 V$), showing the characteristic behavior of contact dominated devices (i.e., gate voltage independence and steep non-linearity at low bias). The top left inset is an optical microscope image of one of our devices (the bar corresponds to 100$\mu$m). The bottom right inset shows the differential conductance normalized at the zero-bias value for 26 different samples, illustrating the reproducibility of the measurements.}
\label{Fig1}}
\end{figure}
The rubrene transistors consist of $\thickapprox 1\mu$m thick rubrene crystals laminated onto doped Silicon substrates (acting as a gate) covered by a 200 nm SiO$_2$ layer, with lithographically defined, electron-beam evaporated Copper electrodes (see Ref.\cite{Ruth} for details). Prior to lamination, the substrate is cleaned in an Oxygen plasma to remove possible resist residues from the dielectric surface. This also influences the Cu contacts by enhancing the oxidation of their surface and possibly changing the material work function. We have therefore performed photo-emission spectroscopy measurements on identically oxidized copper films, and found a value of 5.5 eV (the measurements were performed in air with the Riken Keiki AC-2 system, on films prepared under identical conditions used for the device assembly, for more details see \cite{RefTais}). This suggests that the Fermi level in oxidized copper should align well with the highest occupied molecular orbital of rubrene, also located roughly 5.5 eV below the vacuum level \cite{Mat}.\\
Transport measurements were performed on sufficiently short channel devices, 
in which the channel resistance is negligible with respect to that of the contacts \cite{APL_repr} \cite{Iulian}. In these contact-dominated devices, the I-V characteristics are gate voltage independent and show a steep non-linear increase of the source drain-current at low bias (Fig 1). To understand the mechanisms of charge injection at the interface we studied the temperature and bias dependence of the source-drain current, see (Fig.2). The current decreases with decreasing temperature and if we plot $ln(I(T)/T^2)$ vs $1/T$ - as normally done for Schottky diodes \cite{Sze} - a linear dependence is observed, indicative of thermally activated transport (Fig.3b). The slope corresponds to the activation energy, which is plotted in Fig.3a as a function of bias. Measurements performed on many different devices exhibit similar behavior and the inset of Fig.3a show that the spread in the measured values of activation energy is rather small.\\
\begin{figure}[ht]
\centering
\includegraphics[width=1\columnwidth]{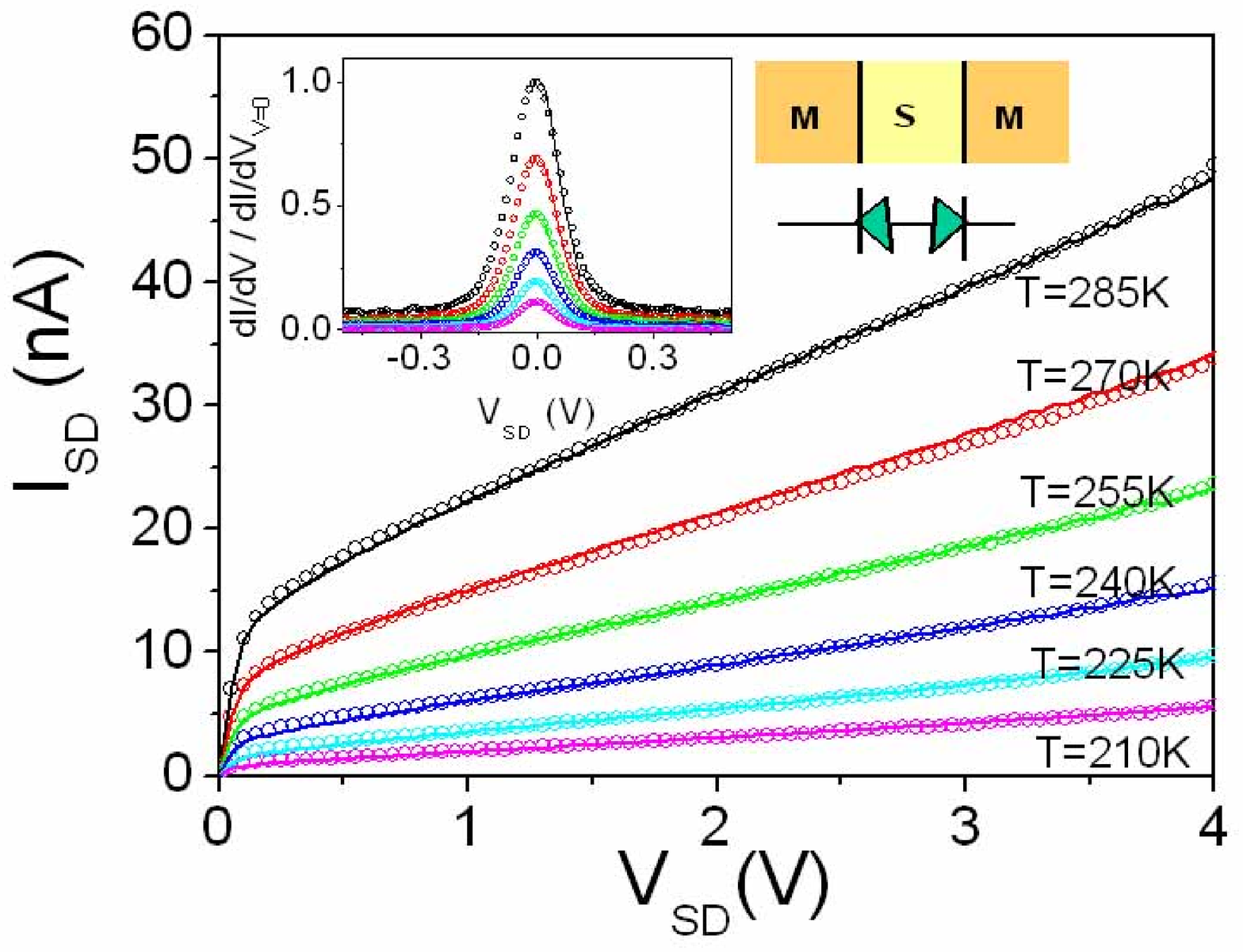}
\noindent{\caption{$I-V$ characteristic of a contact-dominate device, measured at $V_{G}=-30 V$, for different temperatures. The inset shows the temperature dependence of the differential conductance (obtained by numerical differentiation of the $I-V$ curves) normalized to the value at V=0 and T=285K. In both graphs, the open symbols represent the experimental measurements and the continuous lines the theoretical values calculated using two Schottky diodes in series (shown in the scheme on the top right), as described in the text.}
\label{Fig2}}
\end{figure}
By modeling the devices as two oppositely biased Schottky diodes connected in series (see scheme in Fig.2), the measured $I-V$ curves can be directly related to the electrical characteristics of the source and drain contacts. Using this approach, we have previously shown that the usual expression $I(V,T)=I_{0}(e^{eV/nkT}-1)$ for the current through a Schottky diode (n $\simeq$ 1 is the ideality factor), with $I_0$ taken as constant, reproduces the low bias peak in the differential conductance (see inset of Fig.2), but not the behavior of the $I-V$ curves at bias higher than few times $kT$ \cite{APL_repr}. Here, we analyze quantitatively the data by taking into account full dependence of $I_0$ on bias and temperature. To this end we use the conventional theory for Schottky diodes \cite{Sze}, modified to include the effect of a tunnel barrier at the metal/semiconductor interface \cite{Wu}, which in our devices is due to the presence of the CuO$_x$ layer at the metal surface.
Mathematically, the presence of a tunnel barrier at the interface does not modify the functional dependence of the current on bias, and can be included as a prefactor (proportional to the transmission coefficient of the barrier) in the expression of I$_{0}$. Physically, however, the presence of a barrier is important, because it causes a weak coupling between the metal and molecules that, as we discuss below, results in a small surface density of states at the surface of the organic semiconductor.\\
Under reverse bias conditions, $I_0$ depends on $V$ because of the voltage-induced Schottky barrier lowering. This includes the image-charge barrier lowering (the so-called Schottky effect; first bracket in Eq.2) and the field-induced change in the dipole due to electrons occupying surface states \cite{Cowley}-\cite{Lep} (second bracket in Eq.2). The expression for the reverse current then reads \cite{Sze}:
\begin{equation}
I_{R}=\tilde{I} T^{2}e^{-\frac{q\Phi}{kT}}\left(1-e^{-\frac{qV}{nkT}}\right) 
\end{equation}
with
\begin{eqnarray}
\Phi=\Phi_{0}-\left[\frac{q^{3}N_{D}}{8\pi^{2}\epsilon^{3}_{0}\epsilon^{3}_{S}}\left(V_{BI}+V\right)\right]^{\frac{1}{4}}\nonumber\\ -\alpha\left[\frac{2qN_{D}}{\epsilon_{0}\epsilon_{S}}\left(V_{BI}+V\right)\right]^\frac{1}{2}
\end{eqnarray}
$\alpha$, defined in Eq. (4) below, is a parameter that quantifies the effect of surface states (and $q$ is the electron charge). The expression for the forward current reads:
\begin{equation}
I_{F}=\tilde{I}T^{2}e^{-\frac{q\Phi_{0}}{kT}}\left(e^{\frac{qV}{nkT}}-1\right) 
\end{equation}
Eq. 1-3 are valid in the thermionic emission regime, which is expected to be correct for low-transparency contacts, as discussed in Ref. \cite{Wu}. Using these expressions, we calculate numerically the $I-V$ curves through the two diodes connected in series.\\
In order to compare the results of the calculations with the experiments, the values of the parameters in Eq.1-3 need to be fixed. Specifically, $N_{D}$ is the density of dopants (unintentionally present) in the rubrene crystals, recently measured \cite{Taishi} to be $N_{D}\thickapprox1-5$ $10^{20}$ m$^{-3}$. $V_{BI}$ is the built in potential, which is theoretically expected to be smaller than the barrier height \cite{Sze}. The precise value in our devices is not known and in the calculation we fix $V_{BI}=80$ meV (smaller than the measured activation energy). We have checked that small changes in this value do not affect the quality of the comparison between theory and experiments. $\epsilon_{S}=3$ is the dielectric constant of rubrene. $\tilde{I}$ is a constant that determines the absolute magnitude of the current, which is proportional to the junction area and to the transmission probability across the CuO$_x$ layer. It drops out if we confine our analysis to the normalized differential conductance (see inset of Fig.2). The height of the Schottky barrier $\Phi_{0}$ and $\alpha$ are used as fitting parameters. Whereas the value of $\alpha$ is not known a priori, the value of $\Phi_{0}$ can be estimated by extrapolating at zero bias the activation energy measured at high bias (see Fig.3) and cannot be varied much without affecting the comparison between theory and data.\\
\begin{figure}[ht]
\centering
\includegraphics[width=1\columnwidth]{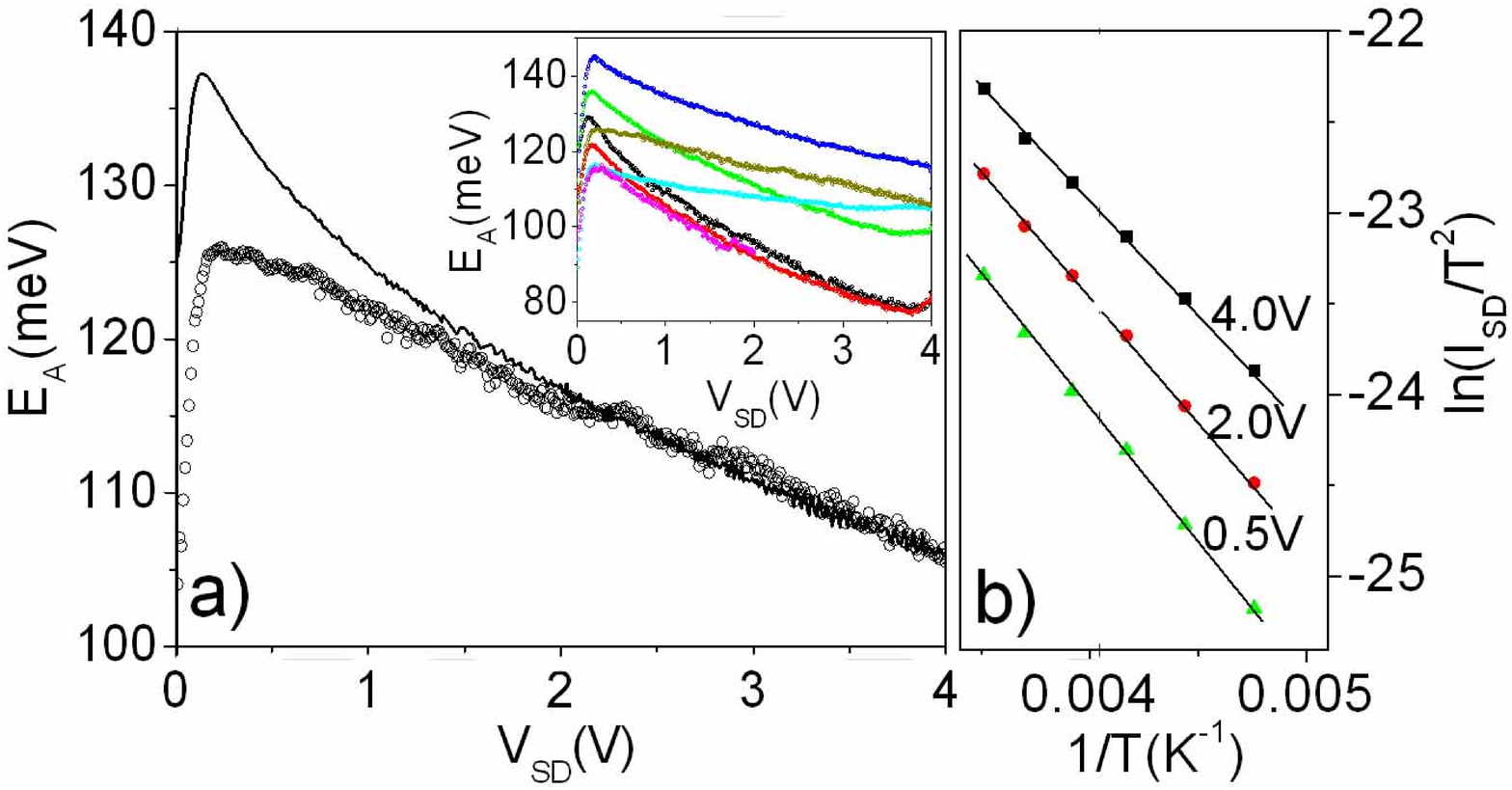}
\noindent{\caption{In (a) the measured activation energy (open dots) is compared with the calculations (continuos line). The inset shows the activation energy measured on several different devices on (SiO$_2$ and Ta$_2$O$_5$), illustrating the relatively small sample-to-sample variation. In (b) the linearity of the plot of ln(I/T) vs 1/T indicates that the thermally activated behavior is experimentally well obeyed for different values of bias voltage (the lines are guides to the eye).}
\label{Fig3}}
\end{figure}
The continuous lines in Fig.2 and 3 show the results of the calculations. Both for the current and for the differential conductance in Fig.2 an excellent 
agreement with the experiments is found. The bias dependence of the activation energy is also reproduced satisfactorily, with only a small deviation at low bias. This quantitative agreement is remarkable, especially if one considers that the theoretical expressions that we have used are strictly speaking valid for a one-dimensional geometry, whereas our transistors are planar devices. The best fits are obtained for $\Phi_{0}$ is $0.13-0.15$ eV and $\alpha \simeq 2$ nm, with the precise values depending on the specific device.\\
In order to interpret the value of $\alpha$ we recall that according to theory \cite{Rhod} \cite{Lep} \cite{Monch}
\begin{equation}
\alpha=\frac{\delta\epsilon_{S}\epsilon_{0}}{\epsilon_{0}+q^{2}\delta D_{S}}
\end{equation}
Here, $D_{S}$ is the density of states at the metal/organic interface (inside the HOMO-LUMO gap). $\delta$ is the thickness of the interfacial dipole layer, which is determined by the decay length of the wave-function of these states inside the bulk of the semiconductor \cite{Bardeen} \cite{Monch}. Its value can be estimated by considering the exponential damping of the wavefunction of an electron under a potential barrier $U$ of approximately 100meV (corresponding to the height of the Schottky barrier). We obtain 1/$\sqrt{\frac{2mU}{\hbar^{2}}}\simeq 6$\AA, where $m$ is the mass of the charge carrier in the HOMO band of rubrene, recently found to be close to the free electron mass \cite{Basov}). Taking this value and the value of $\alpha$ obtained from fitting the data, we use Eq. (4) to estimate the surface density of states. We find $D_{S}\simeq 5$ $10^{11}$ - 1 $10^{12}$ $eV^{-1}cm^{-2}$, between one and two orders of magnitude lower than what is tipically measured in clean metal/organic semiconductor interfaces by means of photo-emission spectroscopy \cite{Vazquez}. \\
The small density of surface states is consistent with the presence of the CuO$_x$ layer at the surface of the copper electrodes. In fact, in an organic semiconductor like rubrene, the mid-gap surface states are induced by hybridization of molecular states due to the coupling to the nearby metal  \cite{Vazquez} (since rubrene molecules are van der Waals bonded, no dangling bonds are present at the crystal surface). If the rubrene molecules are only weakly coupled to the metal -owing to the presence of the oxide layer acting as a tunnel barrier- hybridization is strongly suppressed and so is the density of surface states \cite{Kahn}.\\
Our analysis, therefore, points to the crucial role of the coupling between metal and molecules in determining the I - V characteristic of metal/organic contacts. The coupling has two main effects on transport. On the one hand, a weak coupling tends to increase the contact resistance, because charge carriers have to tunnel through a low transmission barrier. On the other hand, a weak coupling suppresses the formation of surface states and prevents the formation of large dipoles, which can substantially increase the height of the Schottky barrier. The competition between these two mechanisms determines the value of the contact resistance (see also \cite{Biscarini} for beautiful recent experiments qualitatevely consistent with this conclusion).\\
In summary, we have used organic single-crystal transistors to show that the conventional theory for transport through Schottky barriers does describe quantitatively the properties of metal/organic interfaces, at least in the case in which the coupling between metal and molecules is weak. 

We gratefully acknowledge H. Xie for experimental help, and NanoNed and NWO for financial support.

\end{document}